ms-astroph.doc

# RR Lyrae Atmospherics: Wrinkles Old and New.  A Preview[1]

George W. Preston
Carnegie Observatories, 813 Santa Barbara Street, Pasadena, CA 91101
gwp@obs.carnegiescience.edu

ABSTRACT

I report some results of an echelle spectroscopic survey of RR Lyrae stars begun in 2006 that I presented in my Henry Norris Lecture of January 4, 2010. Topics include (1) atmospheric velocity gradients, (2) phase-dependent envelope turbulence as it relates to Peterson's discoveries of axial rotation on the horizontal branch and to Stothers' explanation of the Blazhko effect, (3) the three apparitions of hydrogen emission during a pulsation cycle, (4) the occurrence of He I lines in emission and absorption, (5) detection of He II emission and metallic line-doubling in Blazhko stars, and finally (6) speculation about what helium observations of RR Lyrae stars in omega Centauri might tell us about the putative helium populations and the horizontal branch of that strange globular cluster.

## 1. INTRODUCTION

Otto Struve, my mentor in Berkeley, pushed me toward the RR Lyrae stars in 1957 during a period of confusion about the nature of Baade's (1946) RRab stars near the Galactic Center. He called my attention to the anomalous period distribution of Baade's stars reported by Gaposchkin (1956), and to the challenge issued by Pavlovskaya (1957).  The issue was a matter of some interest at the time, because the short periods reported by Gaposchkin called to mind Kukarkin's (1949) report that RRab stars with periods less than 0.42 d were found preferentially at lower galactic latitudes than those of typical halo stars.  This, coupled with the discovery of two strong-lined RR Lyrae stars by Münch & Terrazas (1946), both at low galactic latitudes and both with short periods, gave rise to suspicion that the Galaxy might contain a second, flattened system of strong-lined RR Lyrae stars, large numbers of which were concentrated toward the Galactic Center.  The conclusion of Morgan and Mayall (1957) that the Bulge luminosity of M31 was dominated by the light of strong-lined giant stars seemed to support this view.  During a coffee break at the corner of Hearst and Euclid in Berkeley Struve set my sail.  "You make something of all this!" he gruffed at me.  I tried, as have many others during the next half century, among them, notably, Hartwick (1987) and Morrison et al. (2009).

After 1965 I wandered off into other topics, returning in 2003 when my interest in RR Lyrae atmospheric phenomena was rekindled by the serendipitous discovery of carbon and s-process rich TY Gruis (Preston et al. 2006), an RRab star with period (0.570 d), almost identical to that of RR Lyrae (0.568 d) itself.  The commonly accepted explanation for such peculiar chemical enrichment of stars in pre-AGB evolutionary states is pollution by the ejecta of an AGB companion (see McClure & Woodsworth 1970).

Wondering how such mass transfer might have affected the HB evolution of TY Gru, I reckoned that further observations were warranted: (1) to see if TY Gru differed from typical RRab stars of similar period in respects additional to its abundance anomalies, and (2) to detect the orbital motion produced by its putative AGB relic companion. Because the antecedent of any AGB companion must have survived RGB evolution intact, the orbital period must be long ($\gtrsim 1$ year). Therefore, the search was likely to be an exercise in patience. To save telescope time I hoped to derive center-of-mass (CoM) velocities from single RV observations, applying phase-dependent RV corrections obtained by use of a standard RV curve. I planned to construct this curve from observations of 10 RRab stars with periods similar to that of TY Gru. Chosen additionally to be bright and southern, exactly half of these turned out to be Blazhko stars (Blazhko 1907), a circumstance that may be viewed as a blessing or a curse. These ten stars are listed in Table 1.

---

[1] This paper preserves the substance and style of remarks that accompanied my PowerPoint presentation of the 2009 Henry Norris Russell lecture

Table 1. Ten RRab stars with pulsation periods similar to that of TY Gru (0.570065 d)

| Star | P(d) | [Fe/H] | V(max. light) | Note |
|---|---|---|---|---|
| UV Oct | 0.54260 | -1.61 | 9.2 | Blazhko |
| AS Vir | 0.55335 | -1.49 | 11.7 | Blazhko |
| V1645 Sgr | 0.55370 | -1.74 | 11.0 | Blazhko |
| DT Hya | 0.56798 | -- | 12.5 | |
| RV Oct | 0.57116 | -1.34 | 10.5 | |
| CD Vel | 0.57349 | -- | 11.7 | Blazhko |
| WY Ant | 0.57433 | -1.66 | 10.4 | |
| BS Aps | 0.58256 | -1.33 | 11.9 | Blazhko |
| Z Mic | 0.58693 | -1.28 | 11.3 | |
| XZ Aps | 0.58724 | -1.57 | 11.9 | |

The goals of the preceding paragraph were not achieved. I could not construct a satisfactory standard RV curve from data for the stars listed in Table 1. This outcome was of no consequence, because TY Gru itself turned out to be a Blazhko variable, a condition that vitiates all attempts to convert single RVs to CoM velocities. The alternative, intensive RV observation annually, produced the results in Figure 1, where horizontal lines at +20 km s$^{-1}$ and -47 km s$^{-1}$ denote arbitrarily drawn upper and lower bounds of the RV data. The CoM velocity appears to have remained constant for the past six years. Thus, as of this writing TY Gru remains on my list of 3carbon and s-process enriched main-sequence stars that fail to show clear evidence of binary motion (Preston 2009a).

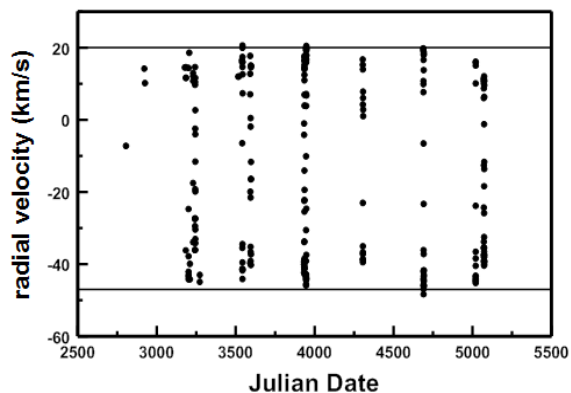

Fig. 1.– Radial velocities of TY Gru plotted versus Julian Date show no obvious secular trend.

The atmospheric behaviors of the stars in Table 1 proved to be so interesting that they themselves became the primary focus of my attention. The remainder of this lecture is devoted to a description of these behaviors, including Blazhko phenomena, to wit: the cyclic, frequently dramatic, modulations of pulsation characteristics that occur in ~ 1/3 of the RRab stars on time scales of 10s to 100s of days (Kolenberg 2008).

All of my observations described herein were made with the 1.5 arcsec entrance aperture of the echelle spectrograph of the du Pont 2.5-m telescope at Las Campanas Observatory. They cover the spectral range $\lambda\lambda 3500 - 9000$ Å with resolution R ~ 27000 at $\lambda 5000$ Å. Time resolution of 3 to 10 min with S/N $\gtrsim$ 20 at $\lambda 4300$ Å is achieved for most of the observations. Spectra for measurement and analysis are created by use of IRAF reduction packages for bias subtraction, flat field division, one-dimensional extraction, and wavelength calibration provided by frequent observation (at least one per hour at each star position) of a Thorium-Argon lamp. Most of what I shall describe is work in progress, so this lecture is properly billed as a preview.

## 2. VELOCITY GRADIENTS IN RRab ATMOSPHERES

During the infall phases of RR Lyrae pulsation the atmospheric layers in which the Balmer lines $H\gamma$, $H\beta$, and $H\alpha$ are formed come to rest progressively later and move progressively more rapidly inward than the layer that produces the metallic-line spectrum (Sanford 1949), evidence for an outward-moving compression/shock wave (van Hoof & Struve 1953) in an atmosphere with a strong velocity gradient that reverses sign during rising light. Hydrodynamic models of RR Lyrae pulsation (Fokin and Gillet 1997) predict complex behaviors throughout a pulsation cycle that depend on fundamental stellar parameters, and they suggest possible points of contact between theory and observation: like the strengths of compression/shock waves, the phases and durations of phases when these waves pass through optically thin layers, and line contours produced by these events.

Examination of Figure 2 suggests how an echelle survey provides such points of contact. The Figure contains measured radial velocities for the metal lines, $H\gamma$, $H\beta$, and $H\alpha$ for two stable RRab stars, Z Mic and RV Oct of nearly identical period P ~ 0.58 d, but substantially different radial velocity amplitudes (ΔRV). Measurements of apparent, rapidly changing violet- and red-shifted absorption features at Hγ, Hβ, and Hα near phase 0.92 for RV Oct are spurious consequences of transient emission. I include them to illustrate measurement difficulties introduced by such emission.

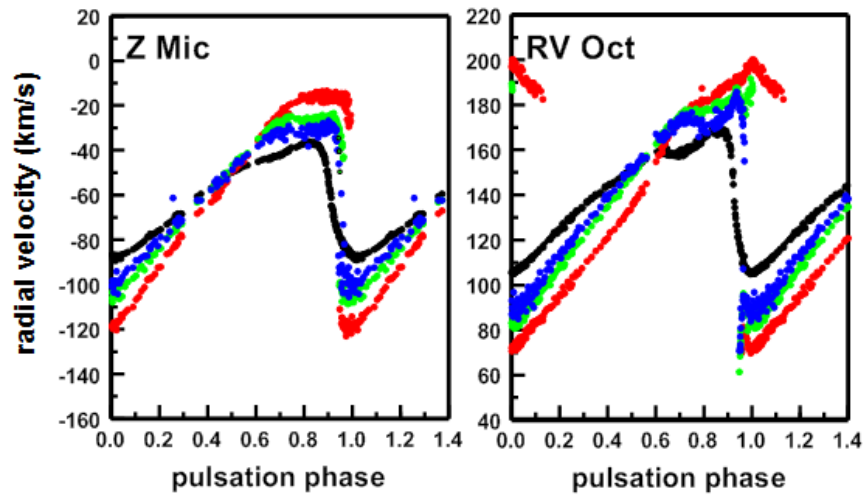

Fig. 2. – Radial velocity data for Z Mic (left) and RV Oct (right), two stable RRab stars in Table 1. Black, blue, green, and red data points denote RVs for metals, $H\gamma$, $H\beta$, and $H\alpha$, respectively.

Three features of these radial velocity curves warrant comment. First, recalling that the sequence metals→Hγ→Hβ→Hα, is one of increasing optical depth at line centers, hence increasing height of formation in the atmosphere, the progression of velocity amplitudes for this sequence in all RRab stars shows that pronounced velocity gradients near the phases of rising light are a general feature of RRab pulsation. Nothing new here. Second, the gradients vanish and reverse sign *at different phases* during declining light, near phase 0.48 for Z Mic, later than phase 0.60 for RV Oct. In addition to star-to-star differences, this phase varies during Blazhko cycles of individual stars. It coincides approximately with an abrupt change in slope of the metallic-line RV curve, a signal phase in the pioneering analysis of Oke, Giver & Searle (1962). Third, the velocity structures of the atmospheres change during pulsation cycles. Thus, in RV Oct the RVs of Hβ are closer to those of Hα at phase 0.1, but they differ substantially from those of Hα and more nearly coincide with those of the metals at phase 0.5. Finally, Liu (1991) showed that the time-average radial velocities of his RRab stars occur at very nearly the same phase, $0.37 \pm 0.02$. This is true for all of the stable RRab stars in Table 1. Therefore, the phases at which velocity gradients vanish always occur during infall, well after the phase of maximum radius. Whether the values of this phase are correlated with other characteristics of RRab stars is a matter for further investigation.

## 3. WHAT CAN BE LEARNED FROM WIDTHS OF ABSORPTION LINES

### 3.1. The Peterson Conundrum

Peterson's surprising discovery (Peterson, Rood & Crocker 1995 and references herein) that many BHB stars possess measurable rotation has sparked lively, ongoing discussion about how such rotation, contrary to reason and all expectations, could appear during the horizontal-branch evolution of subdwarfs, known to be invariably devoid of measurable rotation. Perhaps this discovery itself should be called Peterson's Conundrum, but in keeping with the content of this paper I chose a more restrictive definition. The puzzle arises from consideration of the line widths of RRab stars. The widths of metallic absorption lines in all RRab stars undergo a characteristic variation during each pulsation cycle like the one for WY Ant shown in Figure 3, produced by the combined effects of the expansion itself (van Hoof & Deurinck 1952), turbulence, shocks (Fokin, Gillet, & Chadid 1999), and axial rotation.  A minimum value of FWHM, which always occurs near the phase of maximum expansion, phase ~ 0.4, sets an upper limit on the projected axial rotation, vsini.  Peterson, Carney, and Latham (1996) found that this minimum value lay at or below their spectral resolution, ~ 10 km s$^{-1}$, for all 29 RRab stars in their sample, "… in stark [sic] contrast to the rotation seen in field blue horizontal branch stars … ."

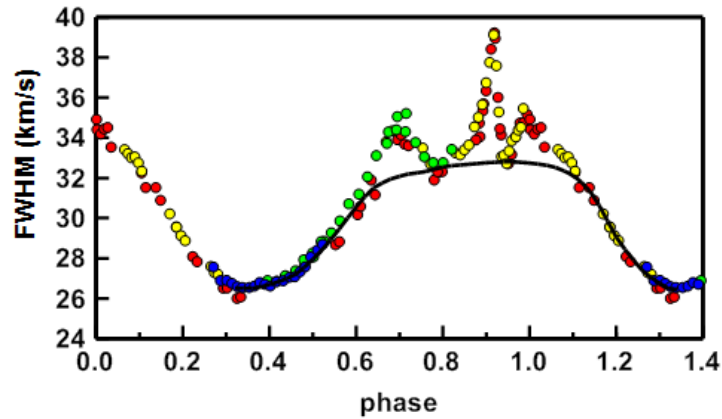

Fig. 3. – The variation of FWHM with pulsation phase for stable WY Ant. The black continuous curve denotes a smooth variation of turbulent broadening on which are superposed three spikes produced by velocity gradients or shocks . Colors distinguish observations made in 2006, 2007, 2008, and 2009.

I explore their mystery in Figure 4, which contains vsini values along the horizontal branch for field stars of Behr (2003) and Kinman et al. (2000) and for several globular clusters in the compilation of Recio-Blanco et al. (2004).  The temperature domain $3.87 < T_{eff} < 4.00$ contains 86 BHB stars, 36 of which (42 %) have vsini > 14.5 km s$^{-1}$.  All of the latter will cross the RR Lyrae gap above the RR Lyrae upper limit and between the two inclined red arrows, trajectories of stars ($v_{rot} \propto T^2$), that conserve angular momentum ($J \propto Rv$) as they evolve toward the AGB at constant luminosity ($L \propto R^2T^4$).  According to this reasoning, and ignoring an unknowable statistical correction required by the upper limits of RRab vsini values, at least 12 of the 29 stars of Peterson et al. (1996) should have vsini > 10 km s$^{-1}$.  None do.  This mismatch between the axial rotations of RRab stars and their adjacent BHB antecedents is the Peterson Conundrum.  Upon completion of my echelle surveys now in progress I will add at least 20 more stars to the RR Lyrae statistic to produce a more robust conclusion.

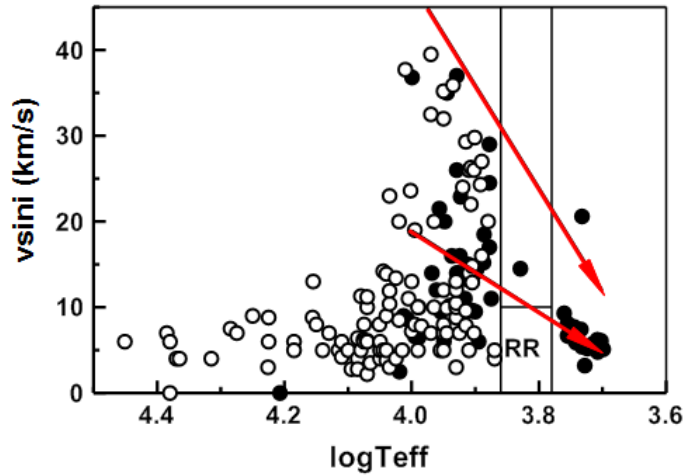

Fig. 4. − Values of vsini for field horizontal branch stars (filled circles) and globular cluster stars (open circles) are plotted versus log $T_{eff}$.  The two vertical lines mark approximate boundaries of the instability strip.  The box at bottom marks the location of 29 RRab stars.  The inclined red arrows are trajectories of stars that conserve angular momentum as they cross the RR Lyrae gap at constant luminosity.

### 3.2. Stothers' Explanation of the Blazhko Effect

Explanations of Blazhko period modulation in terms of the interaction of radial and non-radial modes of similar frequency abound (Kolenberg et al. 2010), but nagging issues have arisen for such models.  Thus RR Lyrae, the archetype of the oblique-dipole rotator model (Balazs-Detre 1964), has no detectable dipole (Chadid et al. 2004).  And in a recent, elegant photometric investigation of another Blazhko star, MV Lyr, Jurcsik et al. (2009) conclude (with thoughtful reservations) that "phase modulations during the Blazhko cycle reflect simply the oscillations of the pulsation period, while the amplitude changes are due to … periodic alterations in the atmospheric structure of the star." This is the view of Stothers (2006, 2010), who hypothesizes that dynamos operating in some RR Lyrae envelopes generate magnetic fields that grow and decay on leisurely (Blazhko) timescales. Magnetic fields so generated suppress turbulent convection, thus producing small changes in the fundamental pulsation period. Suppression of convection is the heart of Stothers' new Blazhko physics, not the dynamo. The latter could be replaced in Stothers' scheme by any cyclic process that alters convective transport of energy through the envelope.

Maybe the variations of FWHM produced by turbulent convection in Blazhko stars, as in Figure 3 above, provide a point of contact between observation and Stothers' hypothesis. To test this possibility I use the FWHM data for UV Oct at the six Blazhko epochs plotted in Figure 5. My

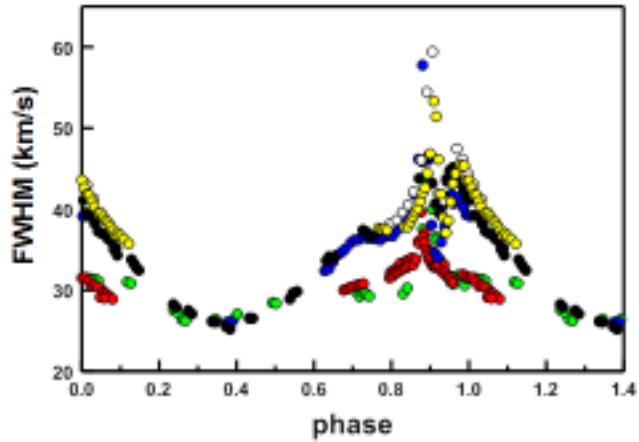

Fig. 5. − Variation of FWHM of metal lines for UV Oct in the years 2006 – 2009. Colors distinguish observations made at different Blazhko phases.

Blazhko period, $P_{Bl}$ = 145 d, derived from variations in $\Delta RV$ and the phase of median RV during rising light agrees well with those based on photometry (Szczygieł & Fabrycky 2007, Guggenberger, Kolenberg, & Medupe 2008).

Minimum values of FWHM appear to be similar in all pulsation cycles of UV Oct, but maximum values differ a lot. Accordingly, I measured FWHM at shock-free locations on 0.7 < phase < 1.1. I used the black continuous curve in Figure 3 as a guide to choose phases for estimation of the turbulence component located between or adjacent to the shock-induced spikes identified by Fokin, Gillet, & Chadid (1999). A plot of these estimates versus Blazhko phase and $\Delta RV$ in Figure 6 shows that the maximum value of turbulent broadening achieved during a pulsation cycle does vary during the Blazhko cycle of UV Oct. However, how this atmospheric turbulence relates to motions in subsurface layers is unknown, and whether this variation of turbulence is of the nature required by Stothers' hypothesis is not yet clear, particularly in view of Stothers' (2010) guarded remarks about the sensitivity of period changes to model temperature. This effort is the first step in a journey to be continued as spectral information for other Blazhko stars is analyzed.

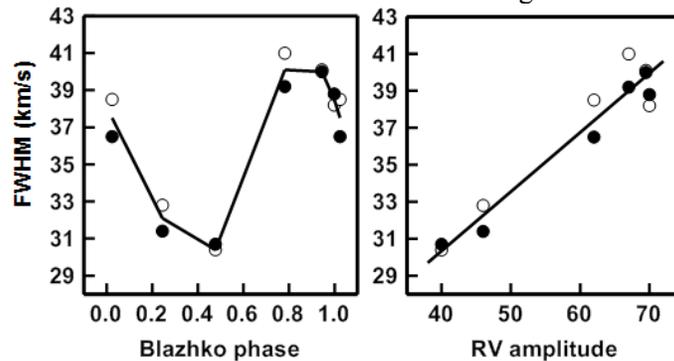

Fig. 6. − FWHM estimated at undisturbed pre-maximum (open circles) and post-maximum (filled circles) pulsation phases for UV Oct plotted versus (left) Blazhko phases, and (right) RV amplitude.

## 4. HYDROGEN LINE PROFILE VARIATIONS

Emission at H$\alpha$ in RR Lyrae during mid-rising light was first reported by Struve & Blaauw (1948), but it was Sanford (1949) who provided graphical illustration of this H$\alpha$ emission profile and the subsequent double absorption lines that prompted Schwarzschild's (1952) shock model. Chronologically, this is the "first apparition". Years later peak H$\gamma$ emission during this first apparition was measured in several RRab stars by Preston & Paczynski (1964). They found no H$\gamma$ emission in the Bailey type b variable SV Eri, understandable in terms of its small RV amplitude, $\Delta RV = 37$ km s$^{-1}$, (Wallerstein 1959: see his Table 1). They also noted, but could not explain, an apparent correlation between emission strength and [Fe/H] evident in their Figure 6, and they suggested that a more ambitious extension of their modest survey might verify what they had found.

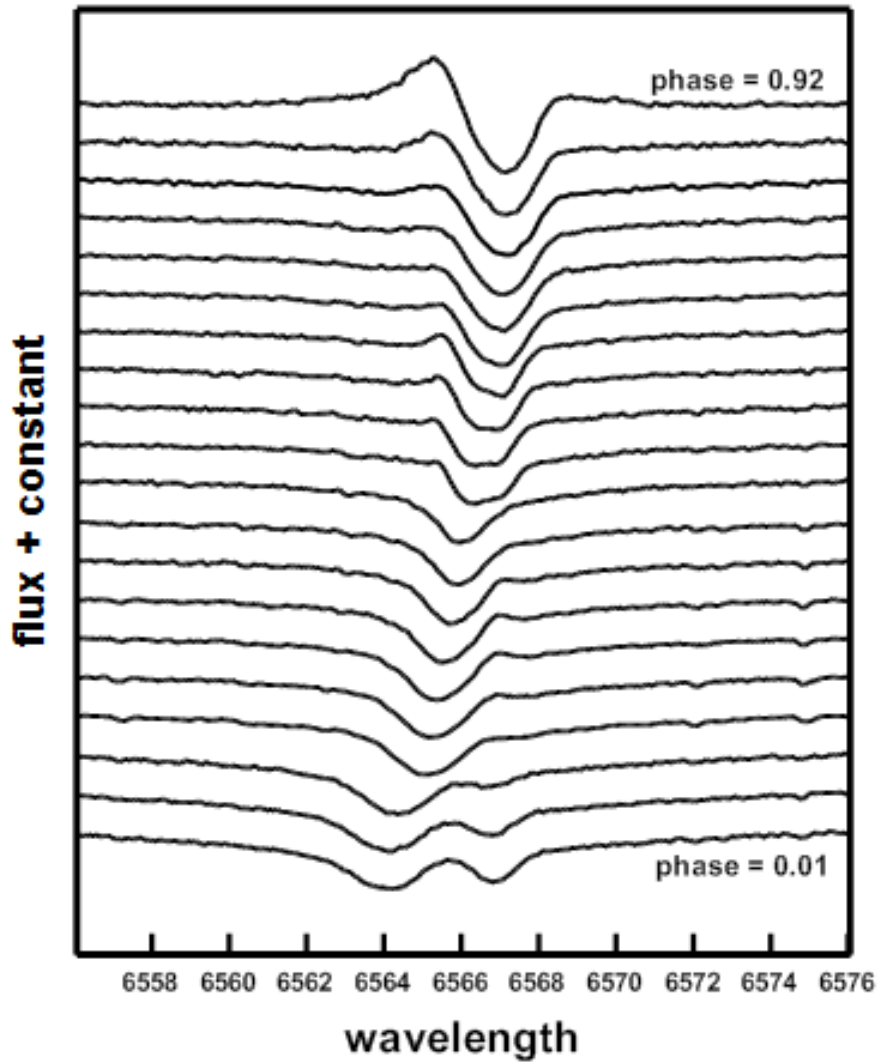

Fig. 7. – The variation of H$\alpha$ profile of RV Oct from maximum light at phase 0.01 (bottom) to the appearance of strong emission at phase 0.92 (top).

Subsequently Gillet and Crowe (1988) discovered the growth and decay of a violet emission shoulder at H$\alpha$ that occurs during the pre-minimum brightening near phase 0.7 in most RRab stars. They attribute this "second apparition" and the modest accompanying increase in luminosity (their "hump") to a shock produced when outer atmospheric layers in "ballistic" infall overtake inner, more slowly collapsing pressure-supported layers. Their discovery illustrates how subtle spectroscopic features lead to improvements in pulsation physics. Now, during the course of the echelle survey described herein, I call attention to yet another emission episode, the "third apparition". These three emission episodes are conveniently viewed in Figure 7, a montage of twenty H$\alpha$ profiles of RV Oct arranged bottom to top in small steps of increasing pulsation phase beginning at 0.01 and ending at 0.92. The zero-point of phase is located at RV minimum, a good surrogate for luminosity maximum (Preston 2009b). The red-shifted, rapidly weakening absorption component in the three bottom spectra is replaced by a weak emission shoulder in the next five spectra – chronologically the "third apparition", the cause of which has not yet been identified. The next two absorption profiles are approximately symmetric. A violet emission shoulder, the "second apparition" of Gillet & Crowe (1988), waxes and wanes in the next six spectra. Note the steady redward shift of the *minimum* within the broad absorption dips in this group of spectra. The boxy profiles in the first two or three of these six spectra, suggest marginal resolution into two components, one replacing the other during declining light. This sequence of profile variations is a common feature of H$\alpha$ line contours in RRab stars during declining light. In the top four spectra the relatively weak and narrow Gillet-Crowe emission is replaced by much broader, stronger emission that appears during primary light rise. In the topmost spectrum at phase 0.92 a red emission wing also appears, reminiscent of the H$\gamma$ profiles of X Ari in Figure 5 of Preston & Paczynski (1964).

Phase durations of the three apparitions of H$\alpha$ emission add up to nearly half of a pulsation cycle; H$\alpha$ emission is hardly the transient once supposed. The evolution of H$\alpha$ profiles during pulsation cycles for WY Ant and XZ Aps, as well as for RV Oct based on many more observations can be viewed as GIF animations in slides 83 – 86 of the PowerPoint file (HNRLecture2009.ppt) for this lecture available in the electronic version of this paper, or at ftp://ftp.obs.carnegiescience.edu/pub/gwp/HNRLecture.

## 5. HELIUM

### 5.1. Ubiquity of He I lines in RRab stars

George Wallerstein (1959) essentially predicted the occurrence of He I emission lines in RR Lyrae stars by the calculations presented in Table 1 of his classic paper about shock waves in Population II variable stars. However, no one bothered to look for George's emission in RRab stars during the next half century. From this we learn that astronomers, at least stellar spectroscopists like me, tend to find what they are looking for.

We now know that sudden onset of He I emission, followed by hour-long subsequent evolution of He I absorption components, particularly D3 ($\lambda$5876 Å) as illustrated in Figure 8, is a general characteristic of RRab stars during rising light (Preston 2009b). Most commonly, a symmetrical emission feature near the CoM velocity appears suddenly at phase 0.92, coincident with strong

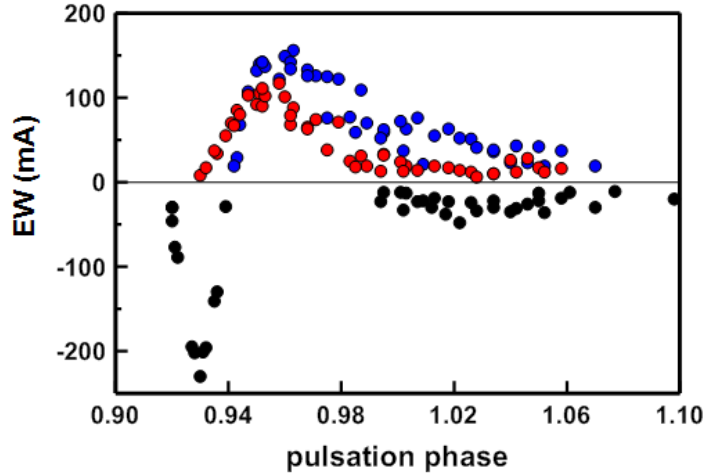

Fig. 8. – The phase variations of He I λ5876 Å line strengths derived from observations of RV Oct during five pulsation cycles in 2006 and 2007: emission equivalent widths (black) are reckoned negative; absorption components (red and violet) are reckoned positive. Note the reappearance of D3 emission after maximum light (phase > 1.0).

Hα emission. At peak emission a few minutes later an absorption feature with velocity of infall suddenly appears, accompanied after a few more minutes by a second absorption component with velocity of outflow, the latter strengthening initially as the infall component reaches maximum strength. Both absorption components then weaken and gradually disappear after maximum light. The persistent reappearance of weak He I emission near maximum light (phase ~ 0.00) is a genuine new wrinkle of RR Lyrae spectroscopy, as is the strongly asymmetric violet wing of the violet-shifted absorption component of D3 shown in Figure 3 of Preston (2009b). I also have detected these phenomena during two cycles of SS Leo and during two large-amplitude cycles of Blazhko UV Oct. However, they are definitely *not* ubiquitous features of RRab pulsation.

Among the stars of Table 1 the He I lines are strongest in RV Oct and weakest in Z Mic. In the former, several He I emission lines are seen (Preston 2009b), while for the latter no emission is seen, only weak absorption at D3 coincident with the appearance of the violet absorption component of Hα. These helium phenomena can be seen in the center panel of Figure 9 below, and they are well-illustrated by the GIF-animations in slides 99 - 101 of PowerPoint file HNRLecture2009.ppt that accompanies the electronic version of this paper, also found at ftp://ftp.obs.carnegiescience.edu/pub/gwp/HNRLecture.

Helium I line emission/absorption sequences occur only during the larger velocity amplitude (ΔRV) cycles of the Blazhko stars, when their behaviors resemble to greater or lesser extent Z Mic (ΔRV = 51 km s$^{-1}$) or RV Oct (ΔRV = 63 km s$^{-1}$). The relation between excitation of H and He and the magnitude of the pulsation velocity implicit in this remark conforms in a general way to expectations of the shock-wave model (Schwarzschild 1952; Wallerstein 1959: see Table 1 and his accompanying text).

Beginning in May 2010 I expanded my echelle survey to encompass the full ranges of period and [Fe/H] encountered among RRab stars. In this enlarged sample (25 stars as of this writing) it is evident that the strengths and durations of He emission and absorptions among the RRab family are related to pulsation period, velocity amplitude, and metal abundance in a complicated way, yet to be analyzed. I illustrate the complexity of the problem by reference to three RRab stars with relatively high [Fe/H]. Neither emission nor absorption has been detected so far in short-period, relatively metal-rich V445 Oph ([Fe/H] = -0.23, P = 0.397 d). However, weak emission and well-marked absorption are seen in longer-period and *more* metal rich AN Ser ([Fe/H] = -0.04, P = 0.522 d), while no emission but well-marked absorption occurs in W Crt ([Fe/H] = -0.50, P = 0.412 d). I use the [Fe/H] values of Layden (1994).

### 5.2. He II Emission and Coincident Line-doubling in Three Blazhko Stars

Weak He II emission at λ4686 Å occurred during phases near peak H and He I emission in AS Vir during the light rise of JD 2454908.7, when the metal-line $\Delta RV$ = 65 km s$^{-1}$, as illustrated in Figure 9. He II λ4686 Å emission of comparable strength occurred in V1645 Sgr during the light rise of JD 2454306.7, when $\Delta RV$ = 64 km s$^{-1}$, and somewhat weaker He II emission occurred in UV Oct during the light rise of JD 2454582.8, when $\Delta RV$ = 70 km s$^{-1}$. These three stars, the only ones in which I have detected He II emission, are all Blazhko stars observed during large-amplitude Blazhko cycles, where "large" here means "larger than the RV amplitude of stable RV Oct", i.e., $\Delta RV$ > 63 km s$^{-1}$. Converted to pulsation velocity (PV) by the projection factor p = 1.35 (Nardetto et al. 2004), the critical amplitude of pulsation velocity for He II emission and line doubling becomes $\Delta PV_{crit}$ > 85 km s$^{-1}$. In all three instances, the He II emission appeared during pulsation cycles in which metallic line-doubling occurred. I illustrate this in Figure 10 by comparing the behavior of the cross-correlation functions (CCF) during rising light returned by the IRAF package fxcor for three Blazhko stars with that of stable RV Oct. Phase increases downward in Figure 10. In the case of AS Vir line-doubling is clearly indicated by two ridgelines that overlap briefly near the phase of He emission. The doubling is less well-marked for V1645 Sgr and only marginally visible as pronounced broadening of the CCF for UV Oct, but the changes in width/structure of the CCF during the He emission phases for the three Blazhko stars clearly differ from the smooth shift of a single ridge line to shorter wavelengths for RV Oct. All observations of these four stars were made with the same spectral resolution (R ~ 25000) and nearly identical time resolutions (a few minutes), so the different behaviors of the CCF are intrinsic to the stars.

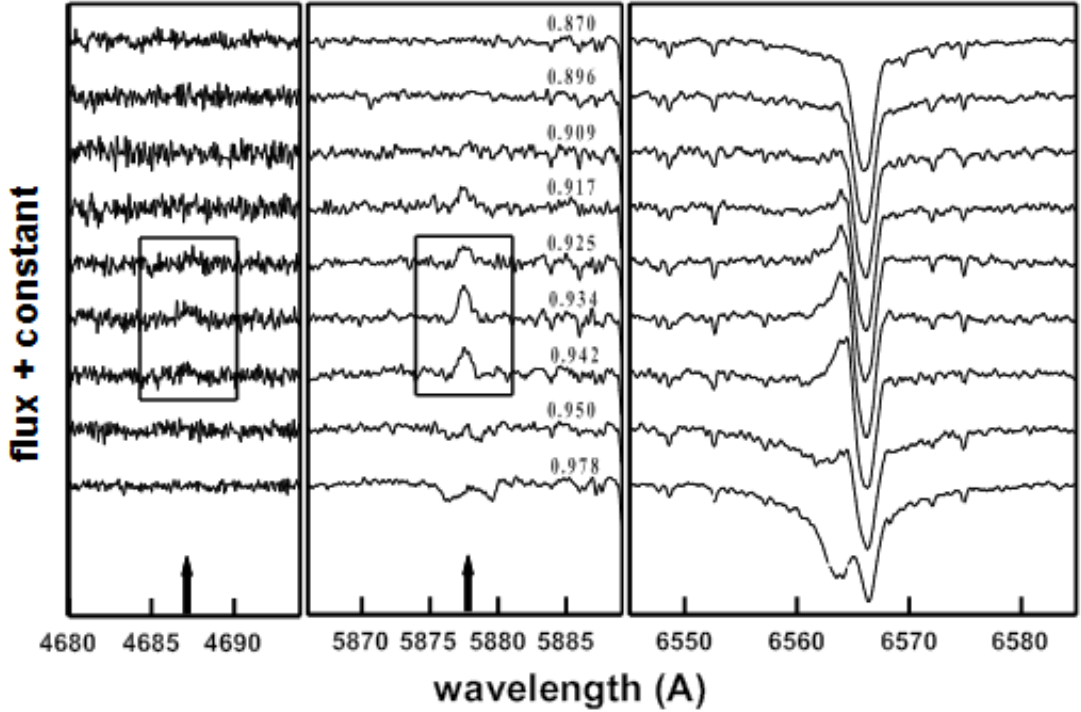

Fig. 9. – Three panels, right to left, show coincidence of peak emission of Hα, D3 He I and λ4686 Å He II for AS Vir. Inset boxes surround He I and He II emission lines in three successive spectra to lead the eye.

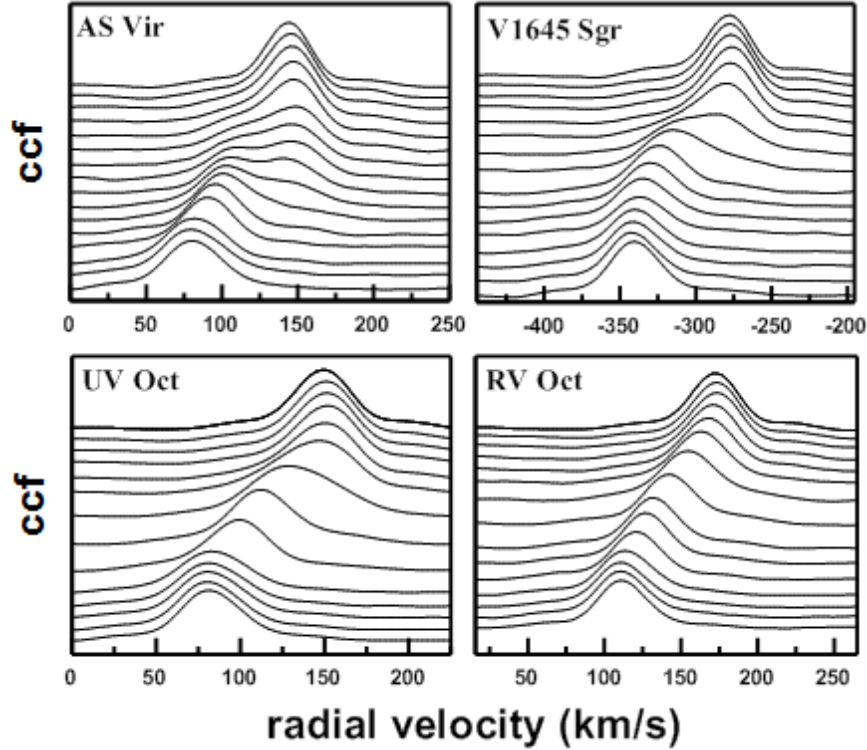

Fig. 10. – Variations of the shapes of cross correlation functions during primary light rises for three Blazhko stars AS Vir, V1645 Sgr, and UV Oct, and for stable RV Oct. Pulsation phase increases downward.

*Considering that metal-line doubling has been reported previously in two additional Blazhko stars, archetype RR Lyr* (Preston et al. 1965, Chadid & Gillet 1996) *and S Ara* (Chadid, Vernin, & Gillet 2008, Chadid et al. 2010), *but not yet reported for any stable RRab star*, it may be appropriate to modify Szeidl's (1987) conclusion that the light amplitudes of Blazhko stars during their large-amplitude cycles lie *on* the period-amplitude relation for stable RRab stars. The helium and line-doubling phenomena described here suggest that the maximum strengths of Blazhko pulsations, however defined, *exceed* those of their stable counterparts. When I examine the period *versus* light amplitude diagram in Szeidl's Figure 3, it seems to me that the Blazhko stars lie systematically above the regression defined by stable RRab stars, in qualitative conformity with the ΔRV > 63 km/s discriminant for line doubling defined in the first paragraph of this section. As noted by Szeidl himself, some or all of the Blazhko stars that appear to lie below the stable ridgeline may not have been observed photometrically during their largest-amplitude Blazhko phases, i.e., they lie low because of observational incompleteness.

## 6. AN APPLICATION: THE HELIUM SEQUENCES OF ω CENTAURI

Globular cluster research has taken a most curious and unexpected turn during the past decade. Not long after Hughes and Wallerstein (2000) reported the existence of an age-metallicity relation among the MSTO stars in ω Centauri, Piotto et al. (2005) reported that the most metal-rich of these stars belong to a second, blue sequence, comprising about 25% of the lower main sequence.

To explain this they must suppose that helium abundance is enhanced by a factor of 1.5 in the stars of the blue sequence. However, it has proven singularly difficult to disentangle the color effects at the turn-off introduced by the possibility of simultaneous variations in metal abundance, helium abundance, and age (Bekki & Norris 2006, Stanford et al. 2006).

Direct measurement of helium would provide comforting assurance about the enhanced-helium interpretation of the blue MS of ω Cen. Recently, Villanova, Piotto, & Gratton (2009) showed that it is possible to derive helium abundances from D3 lines of BHB stars near the blue edge of the RR Lyrae gap in NGC 6752, and Piotto (2009) states that a project to accomplish a similar task is underway for ω Cen as well. However, the D3 lines measured in NGC 6752 are very weak, EW ~ 10 mÅ. The absorption components of this line in metal-poor RR Lyrae stars can be an order-of-magnitude stronger (see Figure 8 above), so perhaps the RR Lyrae stars can help. There are lots of them in ω Cen, 161 according to Clement et al. (2001), and they share the abundance spread (Rey et al. 2000) of their antecedent MSTO and RGB stars (Suntzeff & Kraft 1996, Norris, Freeman, & Mighell 1996). The dominantly blue HB of ω Cen (Rey et al. 2004) guarantees that the bulk of MSTO stars in ω Cen will traverse the RR Lyrae gap on their paths to the AGB, so helium-rich RR Lyraes should be present, if current views about extraordinary helium enrichment in ω Cen are correct. The D3 line contours in the center panel of Figure 9 were obtained during 250 s integrations of an $11^{th}$ mag. star with the echelle spectrograph of a 2.5-m telescope, so the $14^{th}$ mag. RRab stars of ω Cen should pose no problem for 8-m class telescopes.

Lack of appropriate shock models from which to calculate helium abundances from line strengths certainly is a problem, but the distribution of D3 line strengths in a large sample of RRab stars in ω Cen could provide qualitative confirmation or rejection of the enhanced helium hypothesis. I have abiding faith in my theoretical colleagues who work in stellar astrophysics: it will only be a matter of time before He/H abundance ratios are calculated routinely from spectra of RRab stars obtained during rising light.

## 7. CONCLUDING REMARKS

In my presentation of this lecture in Washington, D. C., I ended with comments about "It Takes a Village" (Clinton 2006). I do so here as well. The "child" in Clinton's book, though born into a family, is inevitably raised in a village, where he/she is exposed daily to the knowledge, skills, attitudes, and beliefs of the various inhabitants of that village, all of whom thus contribute to the child's maturation into a successful adult. There is an analogy for this view in the world of science. Nearly all of us inhabit an astronomy department, a physics department, or an observatory. My academic village, Carnegie Observatories, is populated by colleagues who build telescopes for me, build instruments to hang on the telescopes, devise software to run the telescopes and the instruments, and devise software to reduce the outputs of these instruments. Then they teach me how to use all this stuff and, when I do so, they argue with me about my results. I learn from them every day of my life. Mine is a nice village. I want everyone to know how aware I am of my blessings.